\documentclass[aps,prd, onecolumn ,nofootinbib,showpacs,preprintnumbers,tightenlines,11pt]{revtex4-1}
\usepackage{fullpage,epsf,amssymb,amsthm,amsfonts,amsmath,latexsym, graphicx,array,extarrows,mathtools,mathstyle}
\usepackage[normalem]{ulem}

\newcommand{\overbar}[1]{\mkern 1.5mu\overline{\mkern-1.5mu#1\mkern-1.5mu}\mkern 1.5mu}

\let\oldsqrt\sqrt
\def\sqrt{\mathpalette\DHLhksqrt}
\def\DHLhksqrt#1#2{%
\setbox0=\hbox{$#1\oldsqrt{#2\,}$}\dimen0=\ht0
\advance\dimen0-0.2\ht0
\setbox2=\hbox{\vrule height\ht0 depth -\dimen0}
{\box0\lower0.4pt\box2}}

\numberwithin{equation}{section}

\begin{document}

\title{Spectral density constraints in quantum field theory}

\author{Peter Lowdon}
\email{lowdon@physik.uzh.ch}

\affiliation{Physik-Institut, Universit\"at Z\"urich, Winterthurerstrasse 190, 8057 Z\"urich, Switzerland}

\preprint{ZU--TH 06/15}

\begin{abstract}
\noindent
Determining the structure of spectral densities is important for understanding the behaviour of any quantum field theory (QFT). However, the exact calculation of these quantities often requires a full non-perturbative description of the theory, which for physically realistic theories such as quantum chromodynamics (QCD) is currently unknown. Nevertheless, it is possible to infer indirect information about these quantities. In this paper we demonstrate an approach for constraining the form of spectral densities associated with QFT propagators, which involves matching the short distance expansion of the spectral representation with the operator product expansion (OPE) of the propagators. As an application of this procedure we analyse the scalar propagator in $\phi^{4}$-theory and the quark propagator in QCD, and show that constraints are obtained on the spectral densities and the OPE condensates. In particular, it is demonstrated that the perturbative and non-perturbative contributions to the quark condensate in QCD can be decomposed, and that the non-perturbative contributions are related to the structure of the continuum component of the scalar spectral density.
\end{abstract}


\ \\
\ \\

\maketitle

\section{Introduction \label{section1}}

Spectral representations of matrix elements were first investigated by K\"all\'{e}n~\cite{Kallen52} and Lehmann~\cite{Lehmann54}, and then later by~\cite{Jost_Lehmann57} and~\cite{Dyson58} among others. An important consequence of these investigations was the discovery of the \textit{K\"all\'{e}n-Lehmann representation} of the two-point function. For an arbitrary quantum field $\Psi$, this representation relates the two-point function of the field $\langle T\{\Psi(x)\Psi(0)\} \rangle$ to an integral convolution between the free field propagator and some spectral density $\rho$. The integral representation enables one to determine interesting information about the analytic structure of correlation functions, and also has many important applications including the establishment of Goldstone's theorem for relativistic local fields~\cite{Strocchi08}. Another important result in quantum field theory (QFT) is the operator product expansion (OPE). This expansion was first proposed by Wilson~\cite{Wilson69} to describe the behaviour of products (or time-ordered products) of fields in the limit of coinciding space-time arguments. Given the renormalised fields $A(x)$ and $B(y)$, the OPE has the form:
\begin{align}
A(x)B(y) \sim \sum_{i=1}^{n} \widetilde{C}_{i}(x-y) \widetilde{\mathcal{O}}_{i}(y)
\label{ope}
\end{align}   
where $\{\widetilde{\mathcal{O}}_{i}(y) \}_{i=1}^{n}$ is a finite set of renormalised fields, $\widetilde{C}_{i}$ are (possibly singular) coefficient functions, and $\sim$ is understood to imply that an insertion of $A(x)B(y) - \sum_{i=1}^{n} \widetilde{C}_{i}(x-y)\widetilde{\mathcal{O}}_{i}(y)$ into any Green's function will vanish in the (weak) limit $x \rightarrow y$. An important feature of the OPE is that both $\widetilde{\mathcal{O}}_{i}(y)$ and the coefficients $\widetilde{C}_{i}$ depend on an auxiliary parameter $\mu$ called the renormalisation scale. For the purpose of the discussions in this paper we are interested in the structure of two-point functions of certain fields $\Psi$. By using the general form of the OPE outlined in Eq.~(\ref{ope}), these Green's functions can be shown to have the following behaviour in the limit $x \rightarrow 0$:
\begin{align}
\langle T\{\Psi(x)\Psi(0)\} \rangle \sim \sum_{i} C_{i}(x) \langle \mathcal{O}_{i}(0) \rangle
\end{align}      
where $\langle \cdot \rangle$ signifies the vacuum expectation value. The conceptual idea of the OPE is that the series provides an asymptotic decomposition of short and long distance degrees of freedom, which in the case of asymptotically free theories such as quantum chromodynamics (QCD) are partitioned between the Wilson coefficients $C_{i}(x)$ and vacuum condensates $\langle \mathcal{O}_{i}(0) \rangle$ respectively. For general theories though, this decomposition is not necessarily so clear-cut~\cite{Novikov_Shifman_Vainshtein_Zakharov85}. Nevertheless, the OPE has many important applications such as in the construction of factorisation theorems~\cite{Collins11} and the calculation of conformal field theories~\cite{QFT_Maths99}, as well as more applied uses like in the determination of QCD observables such as $R_{had}$~\cite{Chetyrkin_Kuhn94}. \\

\noindent
The spectral representation and the OPE are important results which have led to both successful experimental predictions and important theoretical developments. In particular, over the last few decades the determination of the perturbative and non-perturbative structure of QCD has significantly progressed due to the application of these results. The method that perhaps best epitomises the successful use of both the spectral representation and the OPE is the Shifman-Vainshtein-Zakharov (SVZ) sum rules~\cite{Shifman_Vainshtein_Zakharov79}. By exploiting the analytic structure of certain correlation functions, this approach introduces a parametrised ansatz for the spectral density $\rho$ and uses this to determine mesonic and hadronic parameters in terms of QCD vacuum condensates such as $\langle \overbar{\psi}\psi \rangle$ and $\langle F_{\mu\nu}^{a}F^{a \mu\nu}\rangle$. Given lattice QCD estimates of these condensates, this then allows one to make a prediction for these parameters. The key point here is that it is not possible to exactly calculate the spectral density associated with a correlation function, the reason being that the complete analytic structure of QCD remains unknown. Instead, one has to constrain the form of $\rho$ indirectly. Another example of a method which constrains the form of spectral densities is the so-called \textit{Weinberg sum rules}~\cite{Weinberg67}. These constraints are derived by performing a short distance expansion of the spectral representation of a correlation function, and inferring that certain linear combinations of the spectral densities must vanish if the correlation function in question has a specific singular behaviour.  \\

\noindent
It is clear that constraining the form of the spectral density is very important if one wants to improve understanding of QCD, as well as other QFTs. In the literature this problem has been pursued in a variety of different ways, the SVZ and the Weinberg sum rules being two of the more developed methods. An interesting approach adopted by~\cite{Bernard_Duncan_LoSecco_Weinberg75} is to generalise the Weinberg sum rules by comparing the short distance spectral representation expansion of a correlator with its OPE. Based on which singular terms appear in the OPE, one can then conclude whether certain linear combinations of the spectral density vanish or not. In a similar manner, the authors in~\cite{Novikov_Shifman_Vainshtein_Zakharov85} compare the expression generated by the large momentum propagator expansion in $\phi^{4}$-theory, with the leading singular terms in the OPE, but in this case with the intention of demonstrating the validity of the OPE itself. The success of this comparison approach between the short distance expanded spectral representation and the OPE, suggests that there may well be more information to be gained by performing a full expansion of both expressions, and then matching the resulting terms order by order in $x$. \\

\noindent
The remainder of this paper is structured as follows: in Sec.~\ref{section2} we perform the short distance matching procedure for the scalar propagator in $\phi^{4}$-theory; in Sec.~\ref{section3} we apply the same procedure to the quark propagator in QCD; and finally in Sec.~\ref{section4} we discuss the relevance of our results and the scope for further applications.

\section{Short distance matching in $\phi^{4}$-theory \label{section2}}

In this section the short distance matching procedure outlined at the end of Sec.~\ref{section1} will be applied to the propagator $\langle T\{\phi(x)\phi(0)\} \rangle$ in $\phi^{4}$ scalar field theory. Given the assumption of some standard QFT axioms\footnote{See~\cite{Kallen52, Lehmann54, Jost_Lehmann57, Dyson58} for more details.}, this propagator has the following spectral representation:
\begin{align}
\langle  T\{\phi(x)\phi(0)\} \rangle = \int_{0}^{\infty} ds \ \rho(s) \ i\Delta_{F}(x;s)
\end{align}
where $i\Delta_{F}(x;s)$ is the free bosonic Feynman propagator, and $\rho(s)$ is the spectral density. As with any QFT, renormalisation of the fields is required in order to remove the divergences which arise as a result of the product of fields being ill defined at coincident space-time points. Once this procedure has been performed, the propagator instead satisfies the following renormalised spectral representation~\cite{Schweber61}: 
\begin{align}
\langle  T\{\phi_{R}(x)\phi_{R}(0)\} \rangle = \int_{0}^{\infty} ds \ \rho(s,\mu,g) \ i\Delta_{F}(x;s)
\label{KL_renorm}
\end{align}
where $\phi_{R}$ is the renormalisation of the bare field $\phi$, and the spectral density $\rho$ is now also dependent on the renormalisation scale $\mu$ and coupling $g$. If one now assumes $x$ to be space-like ($x^{2}<0$), the Lorentz invariance of the propagator enables one (for simplicity) to set $x_{0}=0$. Under these conditions the free boson propagator $i\Delta_{F}(x;s)$ has the following exact form~\cite{Huang10}:
\begin{align}
i\Delta_{F}(x_{0}=0,{\bf{x}};s) = \frac{\sqrt{s}}{4\pi^{2}|{\bf{x}}|}K_{1}(\sqrt{s}|{\bf{x}}|)
\label{boson_prop}
\end{align} 
where $K_{1}$ is a modified Bessel function of the second kind. Under the assumption that the small-$|{\bf{x}}|$ behaviour of the integral in Eq.~(\ref{KL_renorm}) can be approximated by expanding the integrand around the point $|{\bf{x}}|=0$, the propagator in this approximation is given by
\begin{align}
\langle  T\{\phi_{R}(x)\phi_{R}(0)\} \rangle &\sim \int_{0}^{\infty} ds \ \rho(s,\mu,g) \Bigg[ \frac{s}{16\pi^{2}}\bigg[ 2\gamma -1 + 2 \ \ln\left(\frac{\sqrt{s}}{2} \right) + 2 \ \ln\left(|{\bf{x}}|\right) \bigg] \nonumber \\
&\hspace{50mm} +\frac{1}{4\pi^{2}|{\bf{x}}|^{2}} + \mathcal{O}(|{\bf{x}}|^{2}) \Bigg] 
\label{spectral_scalar} 
\end{align}
Moreover, the renormalised propagator also has the following OPE~\cite{Collins84}:
\begin{align}
\langle  T\{\phi_{R}(x)\phi_{R}(0)\} \rangle \sim \ & C_{\mathbb{I}}(x, \mu, m, g) + C_{\phi^{2}}(x, \mu, m, g)\langle  \phi_{R}^{2}(0)  \rangle + \cdots
\label{ope_scalar_wilson}
\end{align}
where $m$ is the renormalised mass parameter, $\phi_{R}^{2}= [ \phi^{2} ]_{R}$ is the renormalisation of the bare field $\phi^{2}$, and $\cdots$ represents other possible non-singular terms. Under the assumption that $x$ is space-like (with $x_{0}=0$), this asymptotic expansion is valid in the limit $|{\bf{x}}|\rightarrow 0$. The Wilson coefficients $C_{\mathbb{I}}$ and $C_{\phi^{2}}$ can be calculated perturbatively, and it turns out that to lowest order in perturbation theory they have the following form~\cite{Osborn06}:
\begin{align}
&C_{\mathbb{I}}(x, \mu, m, g) = \frac{1}{4\pi^{2}|{\bf{x}}|^{2}} + \frac{m^{2}}{16\pi^{2}} \ln \left( \mu^{2}|{\bf{x}}|^{2}\right) + \mathcal{O}(g^{2}) \\
&C_{\phi^{2}}(x, \mu, m, g)  = 1+\frac{g}{32\pi^{2}}\ln \left( \mu^{2}|{\bf{x}}|^{2} \right) + \mathcal{O}(g^{2})
\end{align}
Inserting these expressions into Eq.~(\ref{ope_scalar_wilson}) then gives
\begin{align}
\langle  T\{\phi_{R}(x)\phi_{R}(0)\} \rangle &\sim \frac{1}{4\pi^{2}|{\bf{x}}|^{2}} + \frac{m^{2}}{16\pi^{2}} \ln \left( \mu^{2}|{\bf{x}}|^{2}\right)  +\frac{g}{32\pi^{2}}\ln \left( \mu^{2}|{\bf{x}}|^{2} \right)\langle \phi_{R}^{2}(0)  +\langle \phi_{R}^{2}(0)  \rangle + \mathcal{O}(g^{2})                                  
\label{OPE_scalar}
\end{align}
Since Eqs.~(\ref{spectral_scalar}) and (\ref{OPE_scalar}) correspond to equivalent descriptions of the propagator in the small-$|{\bf{x}}|$ limit, and the spectral density $\rho$ is not $x$ dependent, one can equate these two equations and match the coefficients of the various $|{\bf{x}}|$-dependent terms. In doing so, one obtains the following relations between the OPE coefficients and certain moments of the spectral density:
\begin{align}
&\mathcal{O}\left(\tfrac{1}{|{\bf{x}}|^{2}}\right): \int_{0}^{\infty} ds \ \rho(s,\mu,g) = 1 \label{scalar_rel_1} \\
&\mathcal{O}\left(|{\bf{x}}|^{0}\right):   \int_{0}^{\infty} ds \ s\rho(s,\mu,g) \left[  2\gamma -1  + 2 \ \ln\left(\frac{\sqrt{s}}{2} \right) \right] = 2m^{2} \ln \left( \mu\right) +  16\pi^{2}\langle \phi_{R}^{2}(0) \rangle \nonumber \\
& \hspace{95mm} + g\ln \left( \mu \right)\langle \phi_{R}^{2}(0) \rangle + \mathcal{O}(g^{2})   \label{scalar_rel_2} \\
&\mathcal{O}\left(\ln\left(|{\bf{x}}|\right)\right):     \int_{0}^{\infty} ds \ s\rho(s,\mu,g)  = m^{2}  +\frac{g}{2}\langle \phi_{R}^{2}(0)  \rangle  + \mathcal{O}(g^{2})  \label{scalar_rel_3}                                
\end{align}
One thing to notice here is that the relation in Eq.~(\ref{scalar_rel_1}) is exact since there are no other $\mathcal{O}\left(\tfrac{1}{|{\bf{x}}|^{2}}\right)$ terms in Eq.~(\ref{spectral_scalar}), and it is not possible to generate another term with this dependence in Eq.~(\ref{OPE_scalar}) no matter what perturbative order $C_{\phi^{2}}$ and $C_{\mathbb{I}}$ are expanded to. Equations~(\ref{scalar_rel_2})--(\ref{scalar_rel_3}) on the other hand are only perturbatively valid to $\mathcal{O}(g^{2})$, since expanding $C_{\phi^{2}}$ and $C_{\mathbb{I}}$ to higher orders may generate additional constant or $\mathcal{O}\left(\ln\left(|{\bf{x}}|\right)\right)$ terms. By inspecting Eq.~(\ref{scalar_rel_3}), it is clear that this is satisfied if the spectral density is given by 
\begin{align}
\rho(s,\mu,g) = \delta(s- m^{2}) +\frac{g}{2}\langle \phi_{R}^{2}(0)  \rangle A(s) + \mathcal{O}(g^{2}) 
\label{scalar_rho}
\end{align}
where $A(s)$ satisfies the normalisation condition
\begin{align}
\int_{0}^{\infty} ds  \ s A(s) = 1
\end{align}
and also implicitly depends on the renormalisation scale $\mu$. From Eq.~(\ref{scalar_rho}) one can see that the spectral density has several interesting features: there is a Dirac delta term which corresponds to the existence of a state with mass $m$ in the theory; the second term has the structure of a continuum component since it contains an explicit factor of the coupling constant $g$, and is hence a by-product of interactions in the theory; and also the second term is premultiplied by the condensate $\langle \phi_{R}^{2}(0)\rangle$, which suggests that the contribution of this continuum component to the spectral density is moderated by the magnitude of the scalar condensate. \\

\noindent
Inserting this expression for the spectral density into Eq.~(\ref{scalar_rel_2}), and ignoring terms of $\mathcal{O}(g^{2})$ and above, one obtains the relation
\begin{align}
& \frac{g}{2}\langle \phi_{R}^{2}(0) \rangle\left[ 2\gamma  - \ln\left(4\right) -1 + \mathcal{I}  \right] +m^{2} \left[  2\gamma -1  + 2 \ \ln\left(\frac{m}{2} \right) \right] =  2m^{2} \ln \left( \mu\right) +  16\pi^{2}\langle \phi_{R}^{2}(0) \rangle \nonumber \\
& \hspace{105mm} + g\ln \left( \mu \right)\langle \phi_{R}^{2}(0) \rangle + \mathcal{O}(g^{2})
\end{align} 
where $\mathcal{I}$ has the form
\begin{align}
\mathcal{I} = \int_{0}^{\infty} ds  \ s  \ln(s) A(s)
\end{align}
Upon rearrangement this gives
\begin{align}
 \langle \phi_{R}^{2}(0) \rangle &= \left( 1  + \frac{g}{16\pi^{2}}\mathcal{I}' \right)^{-1}\left[ \mathcal{C} - \frac{m^{2}}{8\pi^{2}}\ln\left(\frac{\mu}{m} \right) \right]  + \mathcal{O}(g^{2}) \nonumber \\
&=  \mathcal{C} - \frac{m^{2}}{8\pi^{2}}\ln\left(\frac{\mu}{m} \right) -\frac{g}{16\pi^{2}}\mathcal{I}'\mathcal{C} + \frac{gm^{2}}{128\pi^{4}}\ln\left(\frac{\mu}{m} \right)\mathcal{I}' + \mathcal{O}(g^{2})
\label{match_eqn}
\end{align}
where $\mathcal{I}'$ and $\mathcal{C}$ are
\begin{align}
&\mathcal{I}'=\ln \left( \mu \right) - \frac{1}{2}\left[ 2\gamma  - \ln\left(4\right)-1  + \mathcal{I}   \right] \\
&\mathcal{C} = \frac{m^{2}}{16\pi^{2}} \left[  2\gamma  -\ln\left(4\right) -1 \right]
\end{align}
A significant feature of the expression for the condensate in Eq.~(\ref{match_eqn}) is that it explicitly depends on $\mathcal{I}$, an integral involving the \textit{a priori} unknown continuum contribution to the spectral density $A(s)$. Because this condensate does not receive any non-perturbative contributions~\cite{Novikov_Shifman_Vainshtein_Zakharov85}, it must have exactly the same form as the purely perturbative expression for $\langle \phi_{R}^{2}(0) \rangle$ computed using the renormalisation of the operator $\phi^{2}(0)$. Therefore, by equating these expressions one can obtain information about the continuum component $A(s)$. \\

\noindent
In general, a renormalised operator $\mathcal{O}_{R}^{i}$ satisfies the following renormalisation group equation (RGE)~\cite{Collins84}:
\begin{align}
\mu \frac{d}{d\mu}\mathcal{O}_{R}^{i} = \sum_{j}\gamma_{ij}\mathcal{O}_{R}^{j} 
\label{rge}
\end{align}
where $\gamma_{ij}$ is the anomalous dimension matrix and $\{\mathcal{O}_{R}^{j}\}$ is a finite closed basis of renormalised operators with dimension $\leq \text{dim}(\mathcal{O}_{R}^{i})$. In $\phi^{4}$-theory $\phi_{R}^{2}(0)$ mixes with the identity operator $\mathbb{I}$, but not with $\phi_{R}(0)$~\cite{Collins84}. The RGE for $\phi_{R}^{2}(0)$ therefore has the form
\begin{align}
\mu \frac{d}{d\mu}\phi_{R}^{2}(0) = \sum_{j}\gamma_{\phi^{2}j}\mathcal{O}_{R}^{j} = \gamma_{\phi^{2}\phi^{2}}\phi_{R}^{2}(0) + \gamma_{\phi^{2}\mathbb{I}}\mathbb{I}     
\label{rg_phi2}
\end{align}
By definition, the vacuum expectation value of $\phi_{R}^{2}(0)$ with the perturbative (Fock space) vacuum state vanishes. However, the vacuum expectation value with the physical non-perturbative vacuum state does not necessarily vanish, and this is what $\langle \phi_{R}^{2}(0) \rangle$ corresponds to both in the preceding and proceeding discussions in this section. After inserting both sides of Eq.~(\ref{rg_phi2}) between the physical vacuum state, one obtains the following RGE for $\langle \phi_{R}^{2}(0) \rangle$:
\begin{align}
\left(\mu \frac{d}{d\mu} - \gamma_{\phi^{2}\phi^{2}} \right)  \langle\phi_{R}^{2}(0)\rangle = \left(\mu \frac{\partial}{\partial \mu} + \beta \frac{\partial}{\partial g} + \gamma_{m}m \frac{\partial}{\partial m} - \gamma_{\phi^{2}\phi^{2}} \right) \langle\phi_{R}^{2}(0)\rangle  =  \gamma_{\phi^{2} \mathbb{I}} 
\label{phi2_vev}
\end{align}           
where $\beta$ is the beta function of the theory and $\gamma_{m}$ is the anomalous mass dimension\footnote{Here we use the opposite sign convention to~\cite{Kleinert_Schulte-Frohlinde01} for $\gamma_{m}$.}. At one-loop order one has~\cite{Osborn06,Kleinert_Schulte-Frohlinde01}
\begin{align}
\beta = \frac{3g^{2}}{16\pi^{2}}, \hspace{4mm} \gamma_{\phi^{2}\phi^{2}} = -2\gamma_{m} = -\frac{g}{16\pi^{2}}, \hspace{4mm} \gamma_{\phi^{2} \mathbb{I}} = -\frac{m^{2}}{8\pi^{2}}
\end{align} 
By choosing a mass-independent renormalised operator basis, in this case $\{\mathbb{I}, \phi_{R}^{2}\}$, the anomalous dimensions can in general become mass dependent~\cite{Collins84}, and this is in fact what happens for $\gamma_{\phi^{2} \mathbb{I}}$. Using the method of characteristics, the solution of Eq.~(\ref{phi2_vev}) is equivalent to the solution of the following set of ordinary differential equations:
\begin{align}
&\frac{d \ln \mu}{dt} = 1 \label{ord1} \\
&\frac{dg}{dt} = \beta = \frac{3g^{2}}{16\pi^{2}}  \label{ord2}\\
&\frac{d m}{dt} = \gamma_{m}m = \frac{g m}{32\pi^{2}} \label{ord3}\\
&\frac{d}{dt}\langle\phi_{R}^{2}(0)\rangle = \gamma_{\phi^{2}\phi^{2}}\langle\phi_{R}^{2}(0)\rangle + \gamma_{\phi^{2} \mathbb{I}}  = -\frac{g}{16\pi^{2}}\langle\phi_{R}^{2}(0)\rangle - \frac{m^{2}}{8\pi^{2}} \label{ord4}
\end{align}
However, in order to obtain unique solutions one must first specify a boundary condition for each equation. Since the variable $t$ has no physical significance and only serves to parametrise the characteristic curves along which solutions are defined, one can choose all of the boundary data to be at $t=0$. For Eq.~(\ref{ord1}) the general solution is given by $\ln \mu = t + c_{1}$, so the integration constant has the form $c_{1} =  \ln \mu(t=0)$. Letting $c_{1} =  \ln(\bar{\mu})$, where $\bar{\mu}$ is some physical scale, the condition $t=0$ is equivalent to $\mu = \bar{\mu}$, and so $t = \ln\left(\frac{\mu}{\bar{\mu}}\right)$. With this choice of boundary condition the solutions of Eqs.~(\ref{ord2})--(\ref{ord3}) can be written
\begin{align}
g = \bar{g}\left(1 - \frac{3\bar{g}t}{16\pi^{2}} \right)^{-1} \hspace{5mm} m = \overbar{m}\left(1 - \frac{3\bar{g}t}{16\pi^{2}}\right)^{-\frac{1}{6}}
\label{gm_s}
\end{align}   
where $\bar{g} = g(t=0)$, $\overbar{m} = m(t=0)$, and the solution of $\langle\phi_{R}^{2}(0)\rangle$ has the form
\begin{align}
\langle\phi_{R}^{2}(0)\rangle = \frac{2\overbar{m}^{2}}{\bar{g}}\left(1 - \frac{3\bar{g}t}{16\pi^{2}} \right)^{\frac{2}{3}} -\frac{2\overbar{m}^{2}}{\bar{g}}\left(1 - \frac{3\bar{g}t}{16\pi^{2}} \right)^{\frac{1}{3}} + \overbar{\langle\phi_{R}^{2}(0)\rangle}\left(1 - \frac{3\bar{g}t}{16\pi^{2}}\right)^{\frac{1}{3}}
\label{vev_s} 
\end{align}
with $\overbar{\langle\phi_{R}^{2}(0)\rangle}= \langle\phi_{R}^{2}(0)\rangle(t=0)$. By inverting the expressions in Eq.~(\ref{gm_s}), one can rewrite the solution in Eq.~(\ref{vev_s}) exclusively in terms of the parameters $g$, $m$ and $t = \ln\left(\frac{\mu}{\bar{\mu}}\right)$. Doing so gives
\begin{align}
\langle\phi_{R}^{2}(0)\rangle = \frac{2m^{2}}{g}+ \overbar{\langle\phi_{R}^{2}(0)\rangle}\left[1 + \frac{3g}{16\pi^{2}}\ln\left(\frac{\mu}{\bar{\mu}}\right)   \right]^{-\frac{1}{3}} - \frac{2m^{2}}{g}\left[1 + \frac{3g}{16\pi^{2}}\ln\left(\frac{\mu}{\bar{\mu}}\right)   \right]^{\frac{1}{3}}
\label{poly_g}
\end{align}
Because this perturbative determination of $\langle\phi_{R}^{2}(0)\rangle$ is valid up to one-loop order, the solution is therefore equal to the following expansion of Eq.~(\ref{poly_g}) up to $\mathcal{O}(g)$:
\begin{align}
\langle\phi_{R}^{2}(0)\rangle = \overbar{\langle\phi_{R}^{2}(0)\rangle} - \frac{m^{2}}{8\pi^{2}}\ln\left(\frac{\mu}{\bar{\mu}}\right) - \frac{g}{16\pi^{2}} \ \ln\left(\frac{\mu}{\bar{\mu}}\right)\overbar{\langle\phi_{R}^{2}(0)\rangle} + \frac{gm^{2}}{128\pi^{4}}\left[\ln\left(\frac{\mu}{\bar{\mu}}\right)\right]^{2} + \mathcal{O}(g^{2})
\label{vev_match}
\end{align}
Finally, one can now compare this equation with the expression for $\langle\phi_{R}^{2}(0)\rangle$ [(Eq.~(\ref{match_eqn})] obtained via the spectral density matching conditions in Eqs.~(\ref{scalar_rel_1})--(\ref{scalar_rel_3}). One can clearly see that these expressions have a very similar form. In fact, using the solutions for $g$ and $m$, one can rewrite Eq.~(\ref{match_eqn}) as follows:
\begin{align}
\langle \phi_{R}^{2}(0) \rangle =  \ &\widetilde{\mathcal{C}} - \frac{m^{2}}{8\pi^{2}}\ln\left(\frac{\mu}{\overbar{m}}\right) + \frac{g}{16\pi^{2}}\widetilde{\mathcal{C}}\ \ln\left(\frac{\mu}{\bar{\mu}}\right) + \frac{gm^{2}}{256\pi^{4}}\ln\left(\frac{\mu}{\bar{\mu}}\right) \nonumber \\
& -\frac{g}{16\pi^{2}}\mathcal{I}'\widetilde{\mathcal{C}} + \frac{gm^{2}}{128\pi^{4}}\ln\left(\frac{\mu}{\overbar{m}} \right)\mathcal{I}' + \mathcal{O}(g^{2}) 
\end{align} 
where the constant $\widetilde{\mathcal{C}}$ is defined as
\begin{align}
\widetilde{\mathcal{C}} = \frac{\overbar{m}^{2}}{16\pi^{2}} \left[  2\gamma -\ln\left(4\right) -1 \right]
\end{align}
By demanding that $\mathcal{I}'$ satisfies the following relation
\begin{align}
\mathcal{I}' = \ln\left(\frac{\mu}{\overbar{m}}\right) + \frac{2\ln\left(\frac{\mu}{\overbar{m}}\right)\left[ \gamma -\ln\left(2\right) \right]}{\left[\left[ 2\gamma -\ln\left(4\right) -1 \right] - 2\ln\left(\frac{\mu}{\overbar{m}}\right) \right] } 
\label{I_constr}
\end{align}  
the expression for $\langle \phi_{R}^{2}(0) \rangle$ becomes
\begin{align}
\langle \phi_{R}^{2}(0) \rangle = \ &\widetilde{\mathcal{C}} - \frac{m^{2}}{8\pi^{2}}\ln\left(\frac{\mu}{\overbar{m}}\right) -\frac{g}{16\pi^{2}}\ln\left(\frac{\mu}{\overbar{m}}\right)\widetilde{\mathcal{C}} + \frac{gm^{2}}{128\pi^{4}}\left[\ln\left(\frac{\mu}{\overbar{m}} \right)\right]^{2} + \mathcal{O}(g^{2})
\end{align}
which has exactly the same form as Eq.~(\ref{vev_match}) if one makes the identification 
\begin{align}
\bar{\mu}=\overbar{m}, \hspace{10mm}  \overbar{\langle\phi_{R}^{2}(0)\rangle} = \widetilde{\mathcal{C}}
\label{match_IC}
\end{align}
So equating the short distance matched and RGE derived expressions for $\langle \phi_{R}^{2}(0) \rangle$ has introduced two new constraints: the functional form of the initial conditions in Eqs.~(\ref{ord1})--(\ref{ord4}) is fixed, and hence the form of $\langle \phi_{R}^{2}(0) \rangle$ is completely specified in terms of the free parameters $\overbar{m}$ and $\bar{g}$; and the condition in Eq.~(\ref{I_constr}) implies that $A(s)$ must satisfy
\begin{align}
&\int_{0}^{\infty} ds  \ s \ \ln(s) A(s) = \ln(4\overbar{m}^{2}) - 2\gamma +1  -\frac{4\ln\left(\frac{\mu}{\overbar{m}}\right)\left[ \gamma -\ln\left(2\right)  \right]}{\left[\left[ 2\gamma -\ln\left(4\right) -1 \right] - 2\ln\left(\frac{\mu}{\overbar{m}}\right) \right]  } 
\label{rho_restrict}
\end{align}
and therefore provides an additional constraint on the form of the spectral density $\rho$. \\

\noindent
Although $\phi^{4}$-theory may well not be physically realistic due to its triviality~\cite{Glimm_Jaffe87,Strocchi13}, the discussion in this section demonstrates that the short distance matching procedure provides a way of determining new constraints and qualitative features of the theory, and in particular the spectral density, which contrasts with numerical-based approaches~\cite{Jarrell_Gubernatis96,Burnier_Rothkopf13,Dudal_Oliveira_Silva14}. Moreover, because this procedure is model independent, since it only relies on the existence of an OPE and a spectral representation, it can also equally be applied to physically realistic theories such as QCD, and this is what we pursue in Sec.~\ref{section3}.

\section{Short distance matching in QCD \label{section3}}

The short distance matching procedure that was performed for $\phi^{4}$-theory is equally applicable to QCD, and in this section we focus in particular on analysing the fermionic quark propagator $\langle  T\{\psi(x)\overline{\psi}(0)\} \rangle$ in this way. For this propagator, the spectral density $\rho$ can be decomposed in spinor space as~\cite{Itzykson_Zuber80}
\begin{align}
\rho(s) = \ &\rho_{S}(s)\mathbb{I} + \rho_{PS}(s)\gamma_{5} + \rho_{V}^{\mu}(s)\gamma_{\mu} + \rho_{PV}^{\mu}(s)\gamma_{5}\gamma_{\mu} + \rho_{T}^{\mu\nu}(s)\sigma_{\mu\nu}
\label{fermion_decomp}
\end{align}
where the spinor indices are suppressed. It turns out that the tensor term in Eq.~(\ref{fermion_decomp}) does not contribute, and furthermore if one assumes the absence of parity violation, then $\rho_{PS} = \rho_{PV}=0$. Combining these results, the quark propagator has the following renormalised spectral representation:
\begin{align}
\langle T\{\psi_{R}(x)\overbar{\psi}_{R}(0)\}\rangle &= \int_{0}^{\infty} ds \ \rho_{V}(s,\mu,g)\ iS_{F}(x;s) +  i\Delta_{F}(x;s)\big[ \rho_{S}(s,\mu,g)  -\sqrt{s}\rho_{V}(s,\mu,g) \big]  
\label{fermion_prop}
\end{align}
where $i\Delta_{F}(x;s)$ and $iS_{F}(x;s)$ are the free bosonic and fermionic Feynman propagators respectively, $\rho_{V}^{\mu}(s=p^{2}) := p^{\mu}\rho_{V}(s)$, and $\psi_{R}$, $\overbar{\psi}_{R}$ are the renormalised bare fields. Assuming $x$ is space-like (and setting $x_{0}=0$), one can perform a small-$|{\bf{x}}|$ expansion in an analogous way to Sec.~\ref{section2}. The space-like structure of the free bosonic propagator is given by Eq.~(\ref{boson_prop}), and for the free fermionic propagator it has the form
\begin{align}
iS_{F}(x_{0}=0,{\bf{x}};s) &= \left[\left(i\partial\!\!\!/ + \sqrt{s}\right)i\Delta_{F}(x;s) \right]_{x_{0}=0} \nonumber \\
&=  -i\gamma^{i}x_{i}\bigg[\frac{s \left[K_{0}(\sqrt{s}|{\bf{x}}|) + K_{2}(\sqrt{s}|{\bf{x}}|)   \right]}{8\pi^{2}|{\bf{x}}|^{2}}  + \frac{\sqrt{s}}{4\pi^{2}|{\bf{x}}|^{3}}K_{1}(\sqrt{s}|{\bf{x}}|) \bigg] \nonumber \\
& \hspace{5mm} + \frac{s}{4\pi^{2}|{\bf{x}}|}K_{1}(\sqrt{s}|{\bf{x}}|) 
\end{align}
Finally, inserting the explicit expressions for the free propagators into Eq.~(\ref{fermion_prop}), and expanding around the point $|{\bf{x}}|=0$, one obtains
\begin{align}
\langle T\{\psi_{R}(x)\overbar{\psi}_{R}(0)\}\rangle \sim \int_{0}^{\infty} ds \ &\rho_{V}(s,\mu,g) \Bigg[ -\frac{i{\bf{x}}\!\!\!/}{2\pi^{2}|{\bf{x}}|^{4}}  +\frac{is{\bf{x}}\!\!\!/}{8\pi^{2}|{\bf{x}}|^{2}} + \mathcal{O}(|{\bf{x}}\!\!\!/|) \Bigg] \nonumber \\
& +  \rho_{S}(s,\mu,g) \Bigg[ \frac{s}{16\pi^{2}}\bigg[2 \ \ln\left(\frac{\sqrt{s}}{2} \right) + 2\gamma -1   + 2 \ \ln\left(|{\bf{x}}|\right) \bigg]  \nonumber \\
& \hspace{40mm} +\frac{1}{4\pi^{2}|{\bf{x}}|^{2}} + \mathcal{O}(|{\bf{x}}|^{2}) \Bigg]
\label{spectral_qcd}
\end{align}
In a similar manner to $\phi^{4}$-theory, the quark propagator\footnote{For simplicity we assume here that there is only one flavour of quark, with mass $m$.} has the following operator product expansion~\cite{Yndurain99}:
\begin{align}
\langle T\{\psi_{R}(x)\overbar{\psi}_{R}(0)\}\rangle \sim \ & C_{\mathbb{I}}(x, \mu, m, g) + C_{\overbar{\psi}\psi}(x, \mu, m, g) \langle  \overbar{\psi}\psi(0)  \rangle + \cdots
\label{ope_scalar_quark}
\end{align} 
where $C_{\mathbb{I}}$ and $C_{\overbar{\psi}\psi}$ satisfy the RGEs
\begin{align}
\mu\frac{d}{d\mu}C_{\overbar{\psi}\psi} = -\gamma_{\overbar{\psi}\psi,\overbar{\psi}\psi}C_{\overbar{\psi}\psi} \hspace{10mm} \mu\frac{d}{d\mu}C_{\mathbb{I}} = -      \gamma_{\overbar{\psi}\psi,\mathbb{I}}C_{\overbar{\psi}\psi}
\end{align}  
With this RGE convention for the Wilson coefficients, the anomalous dimensions at one-loop order are given by~\cite{Chetyrkin_Maier10}\footnote{As in Sec.~\ref{section2}, we adopt a mass-independent renormalised operator basis here (like~\cite{Yndurain99}), which means that the anomalous dimensions can in general be mass dependent, unlike in~\cite{Chetyrkin_Maier10}. Nevertheless, the mass-dependent anomalous dimensions are related to the mass-independent ones by a multiplication of a certain power in the mass $m$ (in this case $\gamma_{\overbar{\psi}\psi,\mathbb{I}} = m^{3}\gamma_{\overbar{\psi}\psi,m^{3}}$), and these choices lead to the same RGE for $\overbar{\psi}\psi(0)$.}
\begin{align}
\gamma_{\overbar{\psi}\psi,\overbar{\psi}\psi} = \frac{g^{2}}{2\pi^{2}}, \hspace{5mm} \gamma_{\overbar{\psi}\psi,\mathbb{I}} = \frac{3m^{3}}{2\pi^{2}}\left(1 + \frac{g^{2}}{3\pi^{2}} \right)
\label{wilson_qcd}
\end{align}
Just like in Sec.~\ref{section2}, one can solve these equations using the method of characteristics. In this case, solving these equations (to one-loop order) requires one to solve the following ordinary differential equations:
\begin{align}
&\frac{d \ln \mu}{dt} = 1 \label{ord1_qcd} \\
&\frac{dg}{dt} = \beta = -\frac{7g^{3}}{16\pi^{2}}  \label{ord2_qcd}\\
&\frac{d m}{dt} = \gamma_{m}m = -\frac{g^{2}m}{2\pi^{2}} \label{ord3_qcd}\\
&\frac{d}{dt}C_{\overbar{\psi}\psi} =  -\frac{g^{2}}{2\pi^{2}}C_{\overbar{\psi}\psi} \label{ord4_qcd} \\
&\frac{d}{dt}C_{\mathbb{I}} = -\frac{3m^{3}}{2\pi^{2}}\left(1 + \frac{g^{2}}{3\pi^{2}} \right)C_{\overbar{\psi}\psi} \label{ord5_qcd}
\end{align}   
With the initial conditions $\mu(t=0) = \frac{1}{|{\bf{x}}|}$, $g(t=0)= \bar{g}$, and $m(t=0)= \overbar{m}$, one has
\begin{align}
g^{2} = \bar{g}^{2}\left(1 + \frac{7\bar{g}^{2}t}{8\pi^{2}} \right)^{-1} \hspace{10mm} m = \overbar{m}\left(1 + \frac{7\bar{g}^{2}t}{8\pi^{2}} \right)^{-\frac{4}{7}}
\label{qcd_g_m}
\end{align}
where $t = \ln\left(\mu |{\bf{x}}|\right)$, and the Wilson coefficients have the following form:
\begin{align}
C_{\overbar{\psi}\psi} &= -\left(1 - \frac{7g^{2}t}{8\pi^{2}} \right)^{\frac{4}{7}} \\
C_{\mathbb{I}} &= -\frac{i {\bf{x}}\!\!\!/}{2\pi^{2}|{\bf{x}}|^{4}} + \frac{m}{4\pi^{2}|{\bf{x}}|^{2}} + \frac{i {\bf{x}}\!\!\!/ m^{2}}{8\pi^{2}|{\bf{x}}|^{2}} + \frac{4m^{3}}{3g^{2}}\left[1+ \frac{3g^{2}}{16\pi^{2}}\left(1 - \frac{7g^{2}t}{8\pi^{2}} \right)^{-1} \right]\left(1 - \frac{7g^{2}t}{8\pi^{2}} \right)^{-\frac{5}{7}} \nonumber \\
& \hspace{4mm} - \frac{4m^{3}}{3g^{2}}\left[1+ \frac{g^{2}}{16\pi^{2}}(3+14t)\left(1 - \frac{7g^{2}t}{8\pi^{2}} \right)^{-1} \right] \left(1 - \frac{7g^{2}t}{8\pi^{2}} \right)^{\frac{11}{7}}
\end{align}
Expanding these expressions to $\mathcal{O}(g^{2})$ and inserting them into Eq.~(\ref{ope_scalar_quark}) gives
\begin{align}
\langle T\{\psi_{R}(x)\overbar{\psi}_{R}(0)\}\rangle \sim &-\frac{i{\bf{x}}\!\!\!/}{2\pi^{2}|{\bf{x}}|^{4}} + \frac{m}{4\pi^{2}|{\bf{x}}|^{2}} + \frac{i {\bf{x}}\!\!\!/ m^{2}}{8\pi^{2}|{\bf{x}}|^{2}}  + \frac{3m^{3}}{2\pi^{2}} \ln\left(\mu|{\bf{x}}|\right) + \frac{g^{2}m^{3}}{2\pi^{4}}\ln\left(\mu|{\bf{x}}|\right) \nonumber \\
&+  \frac{3g^{2}m^{3}}{4\pi^{4}}\left[\ln\left(\mu|{\bf{x}}|\right) \right]^{2} - \langle  \overbar{\psi}\psi(0)\rangle  + \frac{g^{2}}{2\pi^{2}}\ln\left(\mu|{\bf{x}}|\right)\langle  \overbar{\psi}\psi(0)\rangle + \mathcal{O}(g^{3})
\end{align}
Since both the spectral densities are $x$ independent, one can perform the same procedure as in Sec.~\ref{section2}, and match the different $|{\bf{x}}|$-dependent coefficients in this expression with the moments of the spectral density in Eq.~\ref{spectral_qcd}:
\begin{align}
&\mathcal{O}\left(\tfrac{{\bf{x}}\!\!\!/}{|{\bf{x}}|^{4}}\right):  \int_{0}^{\infty} ds \ \rho_{V}(s,\mu,g) = 1 \label{qcd_rel_1} \\
&\mathcal{O}\left(\tfrac{1}{|{\bf{x}}|^{2}}\right):  \int_{0}^{\infty} ds \ \rho_{S}(s,\mu,g) = m \label{qcd_rel_2}  \\
&\mathcal{O}\left(\tfrac{{\bf{x}}\!\!\!/}{|{\bf{x}}|^{2}}\right):  \int_{0}^{\infty} ds \ s\rho_{V}(s,\mu,g) = m^{2} \label{qcd_rel_3}  \\
&\mathcal{O}\left(|{\bf{x}}|^{0}\right): \int_{0}^{\infty} ds \ s\rho_{S}(s,\mu,g) \left[  2\gamma -1  + 2 \ \ln\left(\frac{\sqrt{s}}{2} \right) \right]  = 24m^{3} \ln\left(\mu\right)  + \frac{8g^{2}m^{3}}{\pi^{2}}\ln\left(\mu\right) \nonumber \\
&\hspace{92mm} + \frac{12g^{2}m^{3}}{\pi^{2}}\left[\ln\left(\mu\right) \right]^{2} -16\pi^{2}\langle \overbar{\psi}\psi(0)\rangle \nonumber \\
&\hspace{92mm} + 8g^{2}\ln\left(\mu\right)\langle  \overbar{\psi}\psi(0)\rangle + \mathcal{O}(g^{3})   \label{qcd_rel_4} \\
&\mathcal{O}\left(\ln\left(|{\bf{x}}|\right)\right):   \int_{0}^{\infty} ds \ s\rho_{S}(s,\mu,g)  = 12m^{3}  + \frac{4g^{2}m^{3}}{\pi^{2}} + \frac{12g^{2}m^{3}}{\pi^{2}}\ln\left(\mu\right)  + 4g^{2} \langle  \overbar{\psi}\psi(0)\rangle + \mathcal{O}(g^{3}) \label{qcd_rel_5}   
\end{align}            
From Eq.~(\ref{qcd_rel_5}) one can see that this relation is satisfied if the scalar spectral density has the following form: 
\begin{align}
\rho_{S}(s,\mu,g) = \ &\left[12 + \frac{4g^{2}}{\pi^{2}} + \frac{12g^{2}}{\pi^{2}}\ln\left(\mu\right) \right]m \,\delta(s- m^{2}) + 4g^{2} \langle  \overbar{\psi}\psi(0)\rangle B(s) + \mathcal{O}(g^{3}) 
\label{S_rho}
\end{align}
where $B(s)$ satisfies the normalisation constraint
\begin{align}
\int_{0}^{\infty} ds  \ s B(s) = 1
\label{B_constraint}
\end{align}
and also implicitly depends on the renormalisation scale $\mu$. It is interesting to note here that $\rho_{S}$ has the same characteristics as the $\phi^{4}$ spectral density in Eq.~(\ref{scalar_rho}): a Dirac delta term, and a continuum contribution $B(s)$ which has an explicit coupling constant and condensate prefactor. \\

Similarly to Sec.~\ref{section2}, by substituting $\rho_{S}$ into Eq.~(\ref{qcd_rel_4}) one can rearrange to obtain an explicit expression for the quark condensate:
\begin{align}
\langle  \overbar{\psi}\psi(0)\rangle = \ &\mathcal{K} + \frac{3m^{3}}{2\pi^{2}} \ln\left(\frac{\mu}{m}\right) + \frac{g^{2}m^{3}}{2\pi^{4}}\ln\left(\frac{\mu}{m}\right) + \frac{g^{2}}{2\pi^{2}} \mathcal{J}'\mathcal{K} + \frac{3g^{2}m^{3}}{4\pi^{4}} \ln\left(\frac{\mu}{m}\right)\mathcal{J}' \nonumber \\
& + \frac{3g^{2}m^{3}}{4\pi^{4}}\left[\left[\ln\left(\mu\right) \right]^{2} - 2\ln\left(m \right)\ln\left(\mu\right) \right] + \mathcal{K}\left[\frac{g^{2}}{3\pi^{2}} + \frac{g^{2}}{\pi^{2}}\ln\left(\mu\right) \right] +\mathcal{O}(g^{3})
\label{match_vev}
\end{align} 
where $\mathcal{K}$ and $\mathcal{J}'$ are given by
\begin{align}
&\mathcal{J}' = \ln\left(\mu\right) -\frac{1}{2}\left[ 2\gamma  -\ln\left(4 \right) -1  + \mathcal{J} \right] \\
&\mathcal{K} = -\frac{3m^{3}}{4\pi^{2}}\left[  2\gamma -\ln\left(4 \right) -1 \right] 
\end{align}
and $\mathcal{J}$ is defined as
\begin{align}
\mathcal{J} = \int_{0}^{\infty} ds  \ s  \ln\left(s \right) B(s)
\end{align}
In just the same way, this condensate explicitly depends on the unknown continuum component of the spectral density $B(s)$. However, unlike the scalar condensate in $\phi^{4}$-theory, $\langle  \overbar{\psi}\psi(0)\rangle$ contains both perturbative \textit{and} non-perturbative contributions~\cite{Spiridonov_Chetyrkin88}, and as we demonstrate it turns out that the non-perturbative contributions arise due to $B(s)$. To make this more precise, one must first calculate the perturbative contributions to $\langle  \overbar{\psi}\psi(0)\rangle$ [denoted $\langle  \overbar{\psi}\psi(0)\rangle_{\text{P}}$] which originate from the renormalisation of $\overbar{\psi}\psi$. The RGE of $\langle  \overbar{\psi}\psi(0)\rangle_{\text{P}}$ is
\begin{align}
\mu \frac{d}{d\mu}\langle  \overbar{\psi}\psi(0)\rangle_{\text{P}} = \gamma_{\overbar{\psi}\psi,\overbar{\psi}\psi} \langle  \overbar{\psi}\psi(0)\rangle_{\text{P}} +  \gamma_{\overbar{\psi}\psi,\mathbb{I}}
\label{rge_psi}
\end{align}
where $\gamma_{\overbar{\psi}\psi,\overbar{\psi}\psi}$ and $\gamma_{\overbar{\psi}\psi,\mathbb{I}}$ are given in Eq.~(\ref{wilson_qcd}). With the boundary condition $\langle \overbar{\psi}\psi(0)\rangle_{\text{P}}(t=0) = \overbar{\langle  \overbar{\psi}\psi(0)\rangle}_{\text{P}}$, the solution to Eq.~(\ref{rge_psi}) has the form
\begin{align}
\langle \overbar{\psi}\psi(0)\rangle_{\text{P}} = \ &\frac{4m^{3}}{3g^{2}}\left[1+ \frac{3g^{2}}{16\pi^{2}}\left(1 - \frac{7g^{2}t}{8\pi^{2}} \right)^{-1} \right]\left(1 - \frac{7g^{2}t}{8\pi^{2}} \right)^{-\frac{9}{7}}  + \overbar{\langle  \overbar{\psi}\psi(0)\rangle}_{\text{P}}\left(1 - \frac{7g^{2}t}{8\pi^{2}} \right)^{-\frac{4}{7}} \nonumber \\
& - \frac{4m^{3}}{3g^{2}}\left[1+ \frac{g^{2}}{16\pi^{2}}(3+14t)\left(1 - \frac{7g^{2}t}{8\pi^{2}} \right)^{-1} \right]\left(1 - \frac{7g^{2}t}{8\pi^{2}} \right)
\end{align}
Because the anomalous dimensions used in Eq.~(\ref{rge_psi}) are only valid up to $\mathcal{O}(g^{2})$, the perturbative expansion of $\langle \overbar{\psi}\psi(0)\rangle_{\text{P}}$ is also only valid up to this order. Performing this expansion gives
\begin{align}
\langle \overbar{\psi}\psi(0)\rangle_{\text{P}} = \ &\overbar{\langle  \overbar{\psi}\psi(0)\rangle}_{\text{P}} + \frac{3m^{3}}{2\pi^{2}}\ln\left(\frac{\mu}{\bar{\mu}}\right) + \frac{g^{2}m^{3}}{2\pi^{4}}\ln\left(\frac{\mu}{\bar{\mu}}\right)  \nonumber \\
&+ \frac{g^{2}}{2\pi^{2}}\ln\left(\frac{\mu}{\bar{\mu}}\right)\overbar{\langle  \overbar{\psi}\psi(0)\rangle}_{\text{P}} + \frac{3g^{2}m^{3}}{2\pi^{4}}\left[\ln\left(\frac{\mu}{\bar{\mu}}\right)\right]^{2} + \mathcal{O}(g^{3}) 
\label{match_exp}
\end{align}
By also expanding Eq.~(\ref{match_vev}) to $\mathcal{O}(g^{2})$, and comparing this expression with Eq.~(\ref{match_exp}), the quark condensate can be decomposed as follows:
\begin{align}
\langle \overbar{\psi}\psi(0)\rangle = \langle \overbar{\psi}\psi(0)\rangle_{\text{P}}  &+ \frac{g^{2}}{2\pi^{2}}\left[ \widetilde{\mathcal{K}} + \frac{3\overbar{m}^{3}}{2\pi^{2}}\ln\left(\frac{\mu}{\overbar{m}} \right) \right]  \Bigg[ \ln\left(2\overbar{m}\right) -\gamma  +\frac{1}{2}(1 -\mathcal{J}) \nonumber \\
&\hspace{25mm} + \frac{  2\left[\ln\left(\overbar{m}\right) \right]^{2}  - \left[  2\gamma -\ln\left(4 \right) -1 \right] \left[ 3\ln\left(\frac{\mu}{\overbar{m}}\right) - \frac{2}{3} - 2\ln\left(\mu\right)  \right] - 2 }{\left[\left[  2\gamma -\ln\left(4 \right) -1 \right] -2 \ln\left(\frac{\mu}{\overbar{m}}\right)\right]}   \Bigg]
\label{decomp}
\end{align}
where $\bar{\mu}=\overbar{m}$ and $\langle \overbar{\psi}\psi(0)\rangle_{\text{P}}$ has the form of Eq.~(\ref{match_exp}) with 
\begin{align}
\overbar{\langle  \overbar{\psi}\psi(0)\rangle}_{\text{P}} = \widetilde{\mathcal{K}}= -\frac{3\overbar{m}^{3}}{4\pi^{2}}\left[  2\gamma -\ln\left(4 \right) -1 \right]
\end{align}  
Since the first term is purely perturbative, it must be the case that the second term parametrises the non-perturbative contributions to the quark condensate, and in particular, the integral $\mathcal{J}$ involving $B(s)$. This explicit decomposition of the quark condensate into perturbative and non-perturbative contributions has not to our knowledge been established before in the literature, and instead has simply been assumed~\cite{Spiridonov_Chetyrkin88}. Moreover, the direct connection between the non-perturbative contributions and the continuum component of the scalar spectral density $B(s)$ has not been made before. This has interesting applications because it means that if one can estimate the form of $B(s)$ from the integral constraints in Eqs.~(\ref{qcd_rel_2}), (\ref{qcd_rel_4}) and (\ref{B_constraint}), one can use Eq.~(\ref{decomp}) to directly estimate the non-perturbative component of $\langle \overbar{\psi}\psi(0)\rangle$. \\  

\noindent
The analysis in this section has demonstrated that by equating the short distance expansion of the spectral representation of the quark propagator in QCD with its OPE, one can obtain novel information. A nice feature of this method, by contrast to more numerical-based approaches~\cite{Alkofer_Detmold_Fischer_Maris04,Karsch_Kitazawa09,Qin_Chang_Liu_Roberts11}, is that it requires practically no theoretical input other than the form of the Wilson coefficients, and yet from this one is able to derive the qualitative structure of the scalar spectral density $\rho_{S}$, impose integral constraints on both $\rho_{S}$ and $\rho_{V}$, and explicitly decompose the perturbative and non-perturbative contributions to $\langle \overbar{\psi}\psi(0)\rangle$. Moreover, unlike techniques such as the SVZ sum rules, phenomenological approximations such as quark-hadron duality~\cite{Shifman01} are not assumed, which makes this approach process independent and therefore applicable to arbitrary correlators. In principle, this approach could also provide useful input for the SVZ sum rules. A key feature of these sum rules is the requirement to introduce a parametrised form of a spectral density~\cite{Colangelo_Khodjamirian01}, and so information obtained about the structure of this spectral density from the short distance matching procedure could be used to provide additional constraints on the corresponding parameters.

\section{Conclusions \label{section4}}

Spectral densities play a central role in determining the dynamics of a QFT, and yet in many instances it is not possible to calculate these objects exactly. This obstruction arises because the non-perturbative structure of these theories is not well understood. Nevertheless, one can infer information about the form of spectral densities by applying general QFT techniques. In particular, in this paper we have demonstrated that by matching the short distance expansion of the spectral representation of the scalar propagator in $\phi^{4}$-theory and the quark propagator in QCD with their respective OPEs, constraints on both the spectral densities and the OPE condensates arise. On a qualitative level these constraints are interesting because they provide new information about the form of the spectral densities, and specifically the structure of the continuum contribution. In the case of QCD, this information can then be used to explicitly decompose the quark condensate $\langle \overbar{\psi}\psi(0)\rangle$ into perturbative and non-perturbative contributions, and it turns out that the non-perturbative contributions are related to the structure of the continuum component of the scalar spectral density. More directly, these constraints may also provide useful information for procedures such as the SVZ sum rules which rely on constructing a parametrised form of the spectral density of certain correlation functions. A nice feature of this short distance matching approach is that it is completely model independent -- it only relies on the existence of an OPE and a spectral representation. So in principle the analysis applied to the scalar and quark propagators in this paper can equally be applied to other interesting correlators such as the gluon propagator, the vector current correlator $\langle T\left\{ J_{\mu}(x)J_{\nu}(0) \right\} \rangle$, or other more general matrix elements, and this could potentially provide some interesting new insights.

\begin{acknowledgments}
I thank Thomas Gehrmann for useful discussions and input. This work was supported by the Swiss National Science Foundation (SNF) under contract CRSII2\_141847.
\end{acknowledgments}

\bibliographystyle{apsrev4-1}
\bibliography{lowdon_paper_refs}

\end{document}